\documentclass[twocolumn,prc,preprintnumbers,superscriptaddress,floatfix]{revtex4}
\usepackage{dcolumn}
\usepackage{bm}
\usepackage{longtable}
\usepackage{mathrsfs}
\usepackage{graphicx,epsfig,latexsym,amssymb}
\usepackage{multirow,amsmath,array,booktabs,color}
\usepackage[section]{placeins}
\usepackage{soul}

\begin{document}

\title{Probing the $^{12}$C+$^{12}$C fusion reaction via zero-degree spectator measurement in the $^{12}$C($^{14}$N,$\alpha d$)$^{20}$Ne quasi-free reaction }

\author{Xue-Jian Wang}
\affiliation{School of Physics, Anhui University, Hefei 230601,China}
\author{Qun-Gang Wen}
\email[E-mail:]{qungang@ahu.edu.cn}
\affiliation{School of Physics, Anhui University, Hefei 230601,China}
\author{Cheng-Bo Li}
\email[E-mail:]{lichengbo@bjast.ac.cn}
\affiliation{Institute of Radiation Technology, Beijing Academy of Science and Technology, Beijing 100089, China}
\author{Hui-Ming Jia}
\email[E-mail:]{jiahm@cnncmail.cn}
\affiliation{Department of Nuclear Physics, China Institute of Atomic Energy, Beijing 102413, China}
\author{Jian-You Guo}
\email[E-mail:]{jianyou@ahu.edu.cn}
\affiliation{School of Physics, Anhui University, Hefei 230601,China}
\author{Cheng-jian Lin}
\affiliation{Department of Nuclear Physics, China Institute of Atomic Energy, Beijing 102413, China}
\author{Lei Yang}
\affiliation{Department of Nuclear Physics, China Institute of Atomic Energy, Beijing 102413, China}
\author{Feng Yang}
\affiliation{Department of Nuclear Physics, China Institute of Atomic Energy, Beijing 102413, China}
\author{Nan-ru Ma}
\affiliation{Department of Nuclear Physics, China Institute of Atomic Energy, Beijing 102413, China}
\author{Pei-wei Wen}
\affiliation{Department of Nuclear Physics, China Institute of Atomic Energy, Beijing 102413, China}
\author{Tian-peng Luo}
\affiliation{Department of Nuclear Physics, China Institute of Atomic Energy, Beijing 102413, China}
\author{Chang Chang}
\affiliation{Department of Nuclear Physics, China Institute of Atomic Energy, Beijing 102413, China}
\author{Xue-peng Sun}
\affiliation{Department of Nuclear Physics, China Institute of Atomic Energy, Beijing 102413, China}
\author{Hai-rui Duan}
\affiliation{Department of Nuclear Physics, China Institute of Atomic Energy, Beijing 102413, China}
\author{Zhi-jie Huang}
\affiliation{Department of Nuclear Physics, China Institute of Atomic Energy, Beijing 102413, China}
\author{Cheng Yin}
\affiliation{Department of Nuclear Physics, China Institute of Atomic Energy, Beijing 102413, China}
\author{Jiong-he Yang}
\affiliation{Department of Nuclear Physics, China Institute of Atomic Energy, Beijing 102413, China}

\date{\today }

\begin{abstract}
The $^{12}\text{C}+^{12}\text{C}$ fusion reaction is a key physical process in stellar evolution and supernova explosions. It not only determines the late evolutionary fate of massive stars but also directly influences the critical conditions for triggering Type Ia supernovae in accreting white dwarfs. In this work, the Trojan horse method (THM) was employed to investigate the $^{12}\text{C}(^{12}\text{C},\alpha_0)^{20}\text{Ne}$ reaction channel of the $^{12}\text{C}+^{12}\text{C}$ fusion process, using $^{14}\text{N}$ as the Trojan horse nucleus. Telescope detectors were placed at $0^\circ$ and $15^\circ$ to design two experimental configurations covering the forward-angle regions where spectator particles are most likely to emerge. By applying the distorted-wave Born approximation (DWBA), two sets of astrophysical $S^*(E)$ factors for the two-body reaction $^{12}\text{C}(^{12}\text{C},\alpha_{0})^{20}\text{Ne}$ were extracted from the three-body reaction $^{12}\text{C}(^{14}\text{N},\alpha d)^{20}\text{Ne}$ and normalized to existing experimental data. The results show that, limited by the overall experimental resolution, the present study cannot resolve fine resonance structures. Within the astrophysical energy region of $0.5\text{--}2\text{ MeV}$, the extracted $S^{*}(E)$ factor exhibits an increasing trend toward lower energies. The $S^{*}(E)$ factor obtained with the $0^\circ$ configuration shows a flatter trend than that obtained with the $15^\circ$ configuration. Supported by the quasi-free reaction simulation results, the divergence between the two data sets may reflect a combination of experimental acceptance effects, finite detector resolution, and possible differences in the relative contributions of reaction mechanisms. This study provides a systematic examination of the experimental design, quasi-free event selection strategy, and interpretation of the underlying physical mechanisms, serving as a useful reference for understanding the role of the $^{12}\text{C}+^{12}\text{C}$ fusion reaction in astrophysical processes.

\vspace{\baselineskip}
\noindent{\textbf{Keywords}: $^{12}$C+$^{12}$C fusion, $\textit{S*(E)}$ factor, Trojan horse method, Quasi-free}

\end{abstract}

\maketitle


\section{Introduction}
Nuclear reactions in astrophysical environments are fundamental to the synthesis of elements in the natural world and play a pivotal role in unveiling the origin and evolution of the universe ~\cite%
{Bertulani2016,Rolfs1988,Burbidge1957}. In stellar interiors, nuclear reactions mainly occur within the Gamow window, where the thermal motion of charged particles provides sufficient energy to overcome the Coulomb barrier. Carbon burning is one of the key processes in the late stages of stellar evolution, typically following the main-sequence phase. It plays a particularly important role in massive stars ~\cite%
{Becker1981,Spillane2007,Bucher2015}. This process not only dictates the evolutionary pathways of massive stars but also determines the critical ignition conditions for Type Ia supernovae in accreting white dwarf systems.

The typical temperature for carbon burning in stellar environments is about 0.8 GK, corresponding to a center-of-mass energy of E$_{G}$ $\approx$ 1.5 $\pm$ 0.3 MeV. More generally, the temperature range of $0.6-1.2$ GK corresponds to center-of-mass energies of roughly $1-3$ MeV in nuclear astrophysics studies. Carbon burning reaction in star proceeds mainly through the following main decay channels ~\cite%
{Jiang2018,Fruet2020,Morales2024}:
\begin{center}$^{12}$C($^{12}$C,$\alpha$)$^{20}$Ne (Q=4.62MeV)

\vspace{0.8em} $^{12}$C($^{12}$C,p)$^{23}$Na (Q=2.24MeV)

\vspace{0.8em} $^{12}$C($^{12}$C,n)$^{23}$Mg (Q=-2.62MeV)
\end{center}

In the low energy region dominated by the Coulomb barrier, the $^{12}$C+$^{12}$C fusion reaction has long been a central topic in nuclear astrophysics because of its critical role in stellar evolution and nucleosynthesis \cite {Patterson1969,Zhang2020,Tumino2018}. As the energy decreases, the reaction cross section drops steeply, and the strong Coulomb barrier makes direct measurements of the $^{12}$C+$^{12}$C cross section within the Gamow window extremely challenging. This difficulty is further complicated by potential low-energy or sub-threshold resonances, whose energies and strengths remain uncertain but could significantly enhance the cross section ~\cite%
{Cooper2009,Mukhamedzhanov2019,Tan2020,Luo2022}.

Current direct measurements of the $^{12}$C+$^{12}$C reaction reach down to $E_{\mathrm{cm}} \approx 2.1$ MeV~\cite{Jiang2018,Fruet2020,Tan2020,Nan2025}. At lower energies, the astrophysical $S^*(E)$ factor is largely inferred from theoretical extrapolations, which suffer from significant uncertainties due to possible molecular-like resonances~\cite{Taniguchi2021,Taniguchi2024,Tang2022}. To address these limitations, indirect approaches such as the Trojan Horse Method (THM) have been widely employed to provide complementary access to the low-energy reaction dynamics~\cite{Tang2022,Tumino2025,Mukhamedzhanov2014,Li20}. Tumino, Spitaleri \textit{et al}. first applied the THM to investigate the $^{12}$C($^{12}$C,$\alpha$)$^{20}$Ne and $^{12}$C($^{12}$C,p)$^{23}$Na reactions in the energy range of 0.8--2.7 MeV~\cite{Tumino2018}, using $^{14}$N as the Trojan horse nucleus to provide the participant $^{12}$C nucleus and the spectator deuteron. Their measurements revealed several pronounced resonances within the astrophysical energy region, consistent with the theoretical predictions of Cooper~\cite%
{Cooper2009}. These resonances suggest a substantial enhancement of the reaction rate at stellar temperatures.

Subsequently, the theoretical approximations, such as the plane-wave impulse approximation (PWIA), used in the THM analysis of the $^{12}$C+$^{12}$C system have been the subject of further discussion. For instance, Mukhamedzhanov \textit{et al}. proposed a modified evaluation of the extracted $S^{*}(E)$ factor, suggesting a reduced rise at low energies~\cite{Mukhamedzhanov2019}. Meanwhile, further analyses noted that different theoretical treatments could introduce variations in reproducing the high-energy direct measurement data and lead to differing predictions for the spectator particle's angular distribution, particularly regarding the forward-angle emission characteristic of the quasi-free mechanism~\cite{Tumino2025}. These discussions highlight the sensitivity of indirect measurements to theoretical inputs and kinematic constraints, emphasizing the necessity of further experimental verification.

Tan, Boeltzig \textit{et al}. carried out new measurements of this reaction using the SAND (a silicon detector array) system combined with particle-$\gamma$ coincidence techniques. The extracted $\textit{S*(E)}$ factors in the center-of-mass energy range of $2.6-5$ MeV show overall good agreement with existing direct measurement data ~\cite%
{Tan2020,Tan2024}. On the theoretical side, Taniguchi and Kimuraem employed the Antisymmetrized Molecular Dynamics (AMD) model to systematically investigate possible $^{24}$Mg resonance formed in the $^{12}$C+$^{12}$C fusion reaction within the Gamow window. Their calculations predicted several 0$^{+}$ and 2$^{+}$ resonant levels, with energies close to those observed in Tumino$^\prime$s experiment ~\cite%
{Taniguchi2021,Taniguchi2024}.

Meanwhile, Nan, Wang \textit{et al}. indirectly studied $^{24}$Mg resonances relevant to $^{12}$C+$^{12}$C fusion using the thick-target inverse kinematics method with the $^{23}$Na + p ~\cite%
{Nan2025}. Resonance parameters extracted via multichannel R-matrix analysis revealed structures in the astrophysical energy region of $0.5-2$ MeV, with resonance positions consistent with both Tumino$^\prime$s experimental results and the AMD model predictions.

Although direct measurements at low energies are generally more reliable, the THM remains one of the few effective approaches capable of probing below 2 MeV energy region and extracting reaction cross sections~\cite{Tumino2025,A. Tumino2007}. The reliability of the extracted results depends on the proper identification of quasi-free events and the suppression of competing reaction mechanisms through combined particle-identification and kinematic constraints. Consequently, independent experimental verification using different experimental conditions and event-selection strategies is important for assessing the robustness of THM results.

In the THM framework, the quasi-free mechanism requires that spectator particles emerge predominantly at small forward angles. Based on this principle, we selected $^{14}$N($^{12}$C+d) as the Trojan horse nucleus and positioned telescope detectors at 0$^\circ$ and 15$^\circ$ to systematically cover different forward angular regions. Furthermore, to investigate the possible dependence of the extracted observables on different spectator kinematic conditions, the intermediate breakup process $^{12}$C + $^{14}$N $\rightarrow$ $^{12}$C + $^{12}$C + $^{2}$H was introduced as an additional kinematic constraint in the data analysis ~\cite{Wen2008,Wen2011}. This approach has already been successfully applied in previous THM studies of the $^{9}$Be(p,$\alpha$)$^{6}$Li reaction.

This work investigates the two-body reaction $^{12}$C($^{12}$C,$\alpha$)$^{20}$Ne in the astrophysical energy region via the three-body reaction $^{12}$C($^{14}$N,$\alpha d$)$^{20}$Ne. By employing two different experimental configurations, detectors were placed to cover distinct forward angular ranges, while the intermediate breakup process was incorporated into the analysis framework. Through these comparative experimental designs and data analysis strategies, the present study aims to evaluate the influence of spectator kinematic selection on the extracted astrophysical $S^{*}(E)$ factor, offering essential experimental insights for a deeper understanding of the behavior of this reaction under astrophysical conditions.

\section{Trojan horse method}
The THM is an indirect measurement approach based on the quasi-free reaction mechanism, and has been widely applied to study the energy dependence of the astrophysical $\textit{S*(E)}$ factor for two-body reactions of astrophysical interest ~\cite%
{Tumino2025,Romano2006,Spitaleri2011,LaCognata2010,Spitaleri2004,Li2015}. This method provides notable advantages for investigating charged-particle reactions in the stellar energy region, as it effectively overcomes the suppression caused by electron screening and the Coulomb barrier in direct measurements. In contrast to direct methods, THM enables access to extremely low Coulomb barrier energies, thereby allowing the investigation of resonances and even sub-threshold resonances that would otherwise remain inaccessible ~\cite%
{LaCognata2012,Wang2024,Jiang2013,Tang2019,su2025}.

As shown in  Eqs.~(\ref{e1})-(\ref{e2}), a two-body charged-particle reaction of astrophysical interest can be indirectly explored by employing a suitably chosen three-body reaction as a surrogate to probe its reaction mechanism:\begin{equation}
A+x\rightarrow C+c  \label{e1}
\end{equation}%
\begin{equation}
A+a\rightarrow C+c+b  \label{e2}
\end{equation}%

A is the incident particle, a = (x + b) is the Trojan Horse nucleus, x is the participant in the two-body reaction, b is the spectator particle, C and c are the final reaction products. Based on the quasi-free condition of the three-body reaction and the principle of energy conservation, the energy correspondence between the target two-body reaction and the selected three-body reaction can be derived :
\begin{equation}
E^{qf} _{Ax} =E_{Aa}(1-\frac{\mu_{Aa}}{\mu_{Bb}}\frac{\mu^{2}_{bx}}{m^{2}_{x}})-\varepsilon_{a}\label{e3}
\end{equation}%

In Eq.~(\ref{e3}), $\varepsilon_{a}$ is the Trojan nuclei$^\prime$s binding  energy. $\textit{E}_{Ax}$ is the energy interval centered on $\textit{E}_{Ax}^{qf}$: $E_{Ax}=(E_{Ax}^{qf}-E_{cut},E_{Ax}^{qf}+E_{cut})$, $\textit{E}_{cut}$ is the energy cutoff determined by the momentum width of the Fermi motion. Consequently, for a given beam energy, the two-body reaction can be investigated over an energy interval around the quasi-free point ~\cite%
{Li2017,Wen2016,Li2020,Lamia2019}.

The theoretical foundation of the THM is rooted in nuclear reaction models based on the post form distorted-wave Born approximation (DWBA) of the T-matrix element ~\cite%
{Tumino2025,Lei2025}. Under the surface approximation, the cross section of a three-body reaction can be expressed in terms of the S-matrix element of the corresponding two-body reaction. The theoretical framework of THM was first introduced by Baur \textit{et al}. in 1986 ~\cite%
{Baur1986}. Within the quasi-free mechanism of direct reactions and employing the plane-wave impulse approximation (PWIA), the three-body cross section can be factorized into three components associated with the two-body reaction: the kinematical factor $\textit{K}_{F}$, the momentum distribution $|W|^{2}$ of the Trojan horse nucleus, and the two-body cross section under the Half-Off-Energy-Shell (HOES) condition:
\begin{equation}
\frac{d^{3}\sigma}{dE_{Cc}d\Omega_{Bb}d\Omega_{Cc}}=K_{F}|W|^{2}(\frac{d\sigma}{d\Omega})^{HOES}_{Ax\rightarrow Cc}\label{e4}
\end{equation}%

Typel \textit{et al}. subsequently performed a systematic and comprehensive study of the THM theory ~\cite%
{Spitaleri2001,Typel2003,Typel2000}. Starting from the exact three-body reaction T-matrix, they applied the DWBA and, through the surface approximation, established a connection between the three-body reaction cross section and the two-body reaction S-matrix. By further introducing simplifications such as the local momentum approximation, the three-body reaction cross section can be factorized into three components, in a form analogous to that derived under the PWIA. Within this theoretical framework, the extracted two-body reaction cross section naturally includes the Coulomb penetration factor ~\cite%
{Mukhamedzhanov2007,Spitaleri2016,Bertulani2018} :
\begin{equation}
\frac{d^{3}\sigma}{dE_{Cc}d\Omega_{Bb}d\Omega_{Cc}}=K_{F}|W|^{2}\sum_{l}p_{l}(\frac{d\sigma_l}{d\Omega})^{OES}_{Ax\rightarrow Cc}\label{e5}
\end{equation}%

In Eq.~(\ref{e5}), $K_{F}$ denotes the kinematic factor, and $|W|^{2}$ represents the momentum distribution of the spectator inside the Trojan Horse nucleus, given by the squared Fourier transform of the inter-cluster radial wave function. The factor $p_{l}$ is identified as the Coulomb penetration correction factor for angular momentum l ; it is introduced to remove the effect of the Coulomb barrier and thus recover a smoother intrinsic energy dependence.

In the present analysis, the Coulomb penetration factor $p_{l}$ was calculated primarily for the l = 0 partial wave. We also evaluated the possible contribution from higher partial wave and found their influence on the overall shape of the extracted $S^*(E)$ factor to be negligible compared with the dominant l = 0 term. Since the absolute magnitude of the THM $S^*(E)$ factor is normalized to direct experimental data, this approximation does not affect our main conclusions concerning its energy dependence ~\cite{Li2026,Li2026a}.

\begin{figure}[htbp]
\includegraphics[width=8.5cm]{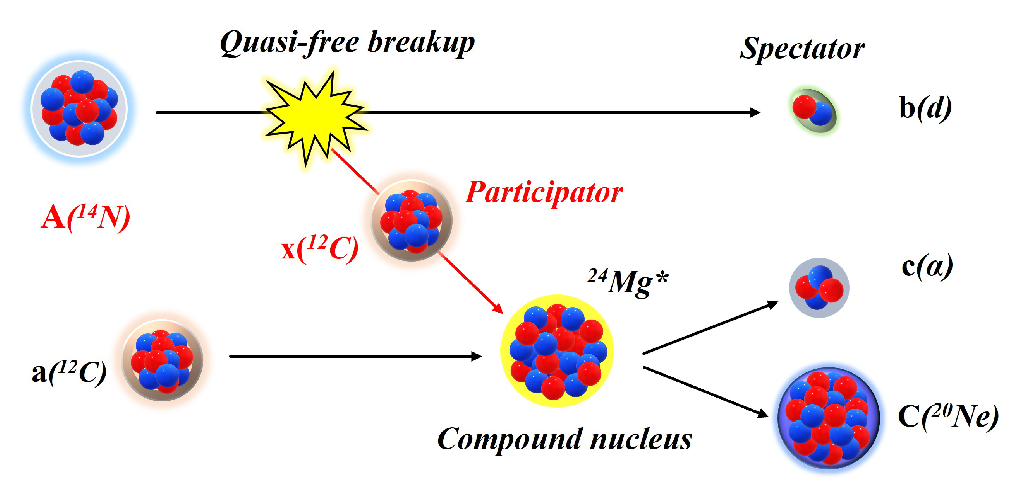}
\caption{Schematic diagram of the Trojan horse method. In the present study, A = (x + b) denotes the incident particle and Trojan Horse nucleus, with x as the participant in the two-body reaction and b as the spectator; C and c denote the final products.}
\label{wi}
\end{figure}

As illustrated in Fig.~\ref{wi}, the present experiment employs $^{14}$N as the Trojan horse nucleus. In the present phenomenological THM treatment, the $^{14}$N nucleus is approximated by a dominant $^{12}$C+d cluster configuration. By selecting appropriate kinematical conditions, the momentum transfer to the d nucleus during the reaction can be minimized and rendered negligible compared with that of $^{12}$C. Under these conditions, the nuclear reaction can be effectively treated as occurring between two $^{12}$C nuclei, while the d nucleus acts as a spectator, essentially retaining the energy and momentum it possessed inside the parent $^{14}$N nucleus. Within this framework, the $^{12}$C nucleus is designated as the participant in the quasi-free reaction, the d nucleus as the spectator, and the parent $^{14}$N nucleus as the Trojan horse nucleus.

\section{Experiment}

The THM requires a kinematically complete measurement of the three-body reaction $^{12}$C + $^{14}$N $\rightarrow$ $\alpha$ +  $^{20}$Ne + d. According to the conservation of energy and momentum, at least two of the three outgoing particles must be detected. In this work, the charged particles $\alpha$ and d were selected to obtain the full kinematic information of the quasi-free process. To achieve precise measurements of the energies and emission angles of the outgoing particles, a $\Delta E-E$ silicon telescope detection system was employed. Each telescope consisted of a 20 $\mu$m single-sided silicon detector (SSSD) and position-sensitive detectors (PSD) with thicknesses of 500 $\mu$m and 1000 $\mu$m. This setup enabled both particle identification and energy determination, allowing accurate extraction of the $\alpha$ and d particle information.

\begin{figure}[htbp]
\includegraphics[width=8.5cm]{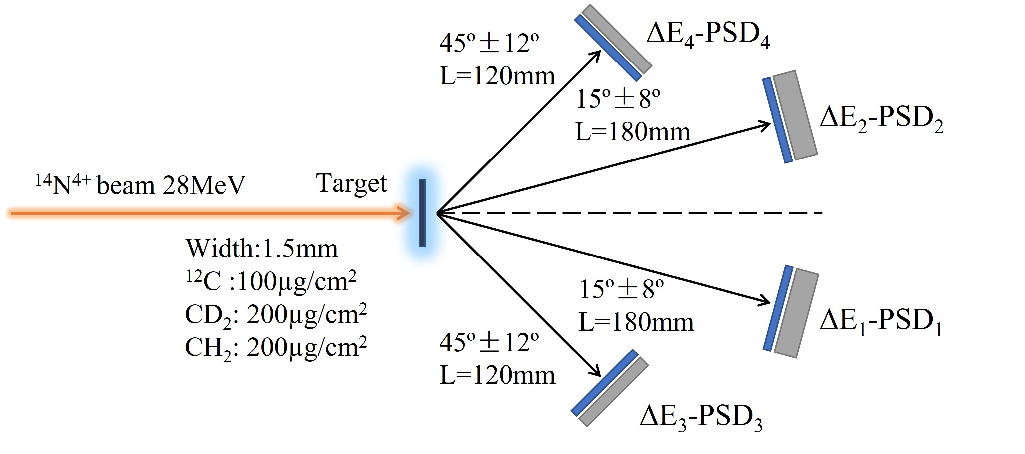}
\caption{Layout and schematic diagram of the detector setup (Configuration 1). The detectors were placed at $\pm$15$^\circ$ and $\pm$45$^\circ$, covering angular ranges of $\pm$(7$^\circ$$\sim$23$^\circ$) and $\pm$(35$^\circ$$\sim$55$^\circ$), respectively, to detect $\alpha$-d coincidence events. This setup is comparable to that employed by Tumino \textit{et al}. (2018) [12].}
\label{wii}
\end{figure}

\begin{figure}[htbp]
\includegraphics[width=8.5cm]{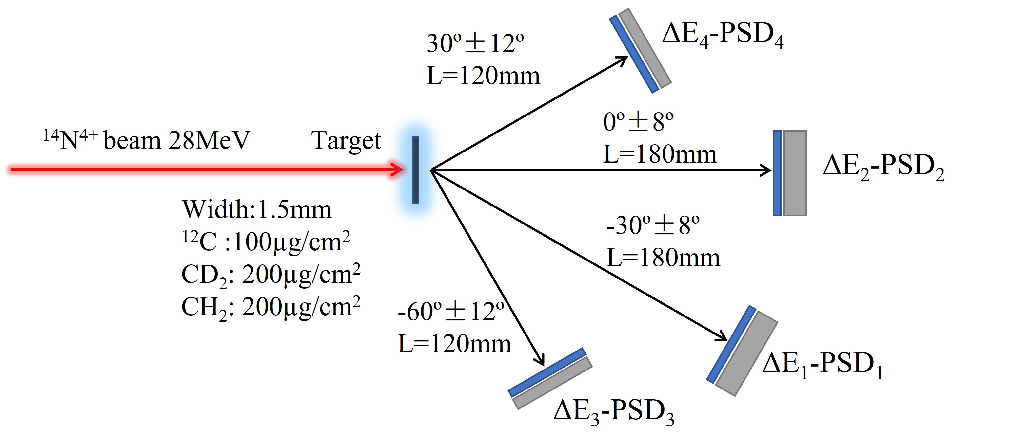}
\caption{Layout and schematic diagram of the detector setup (Configuration 2). The detectors were positioned at 0$^\circ$, $\pm$30$^\circ$, and -60$^\circ$, covering angular ranges of -8$^\circ$$\sim$8$^\circ$,  $\pm$(18$^\circ$$\sim$42$^\circ$) and -48$^\circ$$\sim$-72$^\circ$, to facilitate measurements of the spectator near zero degrees.}
\label{wiii}
\end{figure}

As shown in Figs.~\ref{wii} and \ref{wiii}, two experimental layouts were designed for this study. In the configuration shown in Fig.~\ref{wii}, the telescope detectors were placed at $\pm$15$^\circ$ and $\pm$45$^\circ$, covering angular ranges of $\pm$(7$^\circ$$\sim$23$^\circ$) and $\pm$(35$^\circ$$\sim$55$^\circ$), respectively, to detect the coincidence events of $\alpha$-d. This setup is comparable to that employed by Tumino \textit{et al}.~\cite%
{Tumino2018}. In the configuration shown in Fig.~\ref{wiii}, the detectors were positioned at 0$^\circ$, $\pm$30$^\circ$, and -60$^\circ$, covering angular ranges of -8$^\circ$$\sim$8$^\circ$,  $\pm$(18$^\circ$$\sim$42$^\circ$) and -48$^\circ$$\sim$-72$^\circ$, to facilitate measurements of the spectator particles near zero degrees. Among these, the detectors located at 0$^\circ$ and $\pm$15$^\circ$ in Fig.~\ref{wii} and Fig.~\ref{wiii} were specifically dedicated to the measurement of the spectator deuteron.

This angular configuration was designed to maximize the detection efficiency of quasi-free reaction events. The detector configurations were as follows: Detector 1: 30 $\mu$m Al + 20 $\mu$m SSSD + 1000 $\mu$m PSD; Detector 2: 15 $\mu$m Cu + 20 $\mu$m SSSD + 1000 $\mu$m PSD; Detector 3: 20 $\mu$m SSSD + 500 $\mu$m PSD; Detector 4: 20 $\mu$m SSSD + 500 $\mu$m PSD.

Considering that the quasi-free reaction mechanism requires the spectator deuterons to be concentrated in the forward small angle region, where detectors are particularly susceptible to background interference from the incident beam, beam-stopper foils were installed in front of Detectors 1 and 2. Specifically, cuprum and aluminum foils with thicknesses of 15 $\mu$m and 30 $\mu$m, respectively, were employed to effectively suppress background contributions originating from the beam, particularly from heavy ions such as C and N. These foils still permit the passage of light charged particles, including protons, deuterons, and alpha particles.

The signals recorded by the PSD were directly correlated with particle energies and emission angles. The PSD achieved a position resolution of approximately 0.3 mm and an energy resolution of $\sim$45 keV for a 5.5 MeV $\alpha$ source. For precise measurements, spatial calibration of the PSDs was carried out using an evenly spaced calibration mask, while energy calibration was performed using a standard $\alpha$ radioactive source in combination with the elastic scattering reactions $^{14}$N(p, p)$^{14}$N and $^{14}$N(d, d)$^{14}$N.

The experiment was carried out at the HI-13 tandem accelerator facility of the China Institute of Atomic Energy (CIAE), Beijing. According to Eq. (3), a beam energy of 28 MeV for $^{14}\text{N}^{4+}$ was selected, corresponding to a quasi-free energy of $E^{qf}_{Ax} = 1.5\text{ MeV}$, to effectively cover the Gamow window of $E_{\text{c.m.}} = 1\text{--}2\text{ MeV}$. The beam, with currents ranging from 10 to 40 nA, bombarded a $^{12}\text{C}$ target with a thickness of $100\text{ }\mu\text{g/cm}^2$ and a width of 1.5 mm.

\begin{figure}[htbp]
\includegraphics[width=9cm]{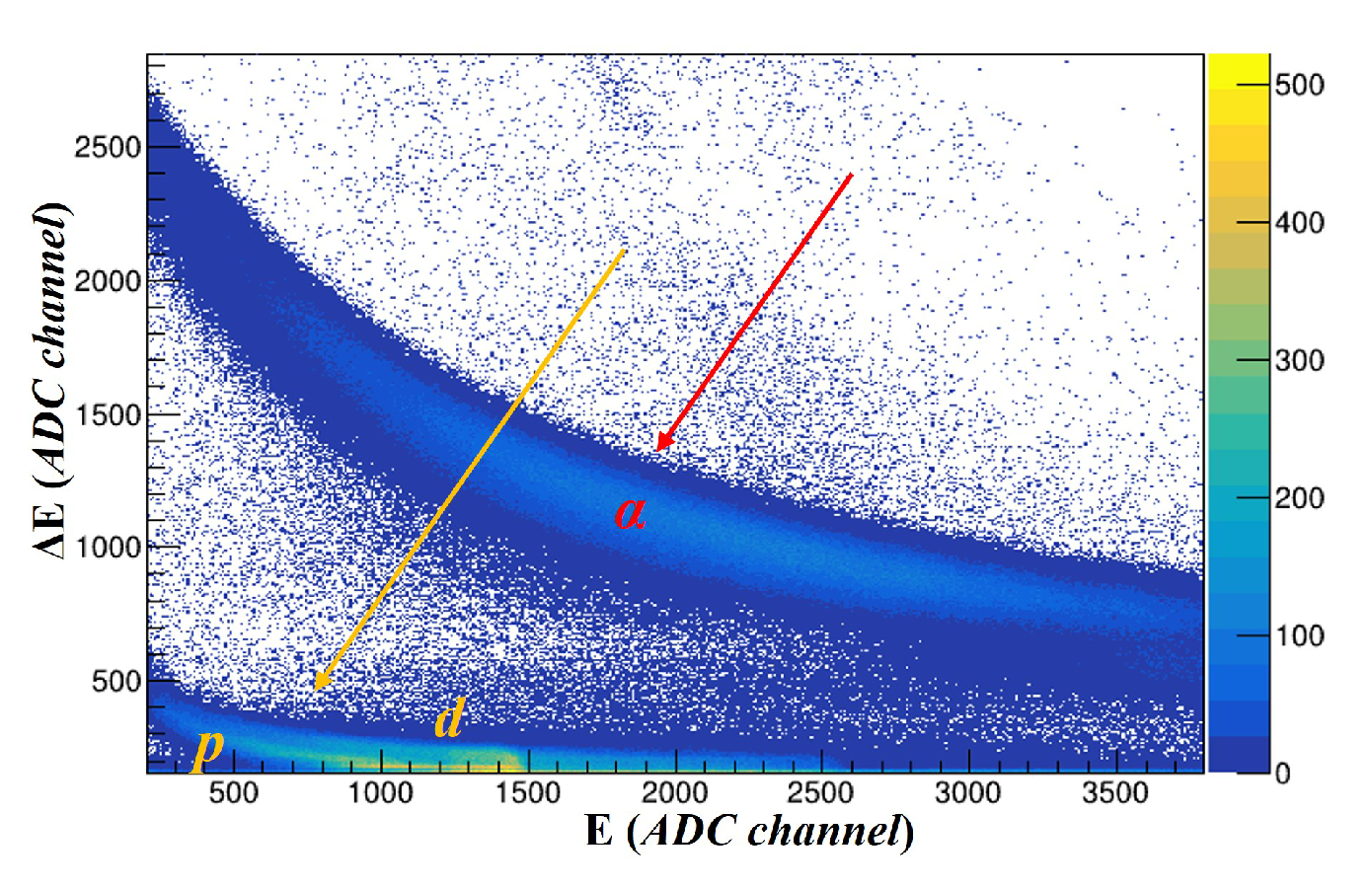}
\caption{Typical $\Delta E-E$ spectrum. Distinct energy loss characteristics of different particles give rise to well-defined energy bands in the two-dimensional $\Delta E-E$ plot.}
\label{wiv}
\end{figure}

\section{DATA analysis}
\subsection{Q-value of the three-body reaction $^{12}$C($^{14}$N,$\alpha d$)$^{20}$Ne}

Fig.~\ref{wiv} presents the $\Delta E-E$ spectrum, where the vertical axis corresponds to the energy deposited in the first-layer SSSD, and the horizontal axis corresponds to the energy deposited in the second-layer PSD. Owing to their distinct energy-loss characteristics, well-defined particle bands appear in the two-dimensional plot. The red arrow marks the alpha-particle band located in the upper-right region, which is well separated from light ions. The yellow arrow indicates the lower-left region where $Z=1$ light charged particles (protons and deuterons) are concentrated.

Although the proton and deuteron loci exhibit partial overlap in the low-energy region, event identification does not rely on graphical $\Delta E-E$ gating alone. Candidate $Z=1$ events are further subjected to multi-tiered kinematic constraints, including strict three-body $Q$-value reconstruction and intermediate breakup selection. Protons originating from competing channels, such as $^{12}\text{C}(^{14}\text{N},\alpha p)^{21}\text{Ne}$, induce a substantial kinematic shift if misidentified as deuterons, causing their reconstructed $Q$-values to fall well outside the target $\alpha_0$ window. Consequently, potential proton contamination is effectively suppressed, minimizing its impact on the target $^{12}\text{C}(^{14}\text{N},\alpha d)^{20}\text{Ne}$ events.

\begin{figure}[htbp]
\includegraphics[width=8.5cm]{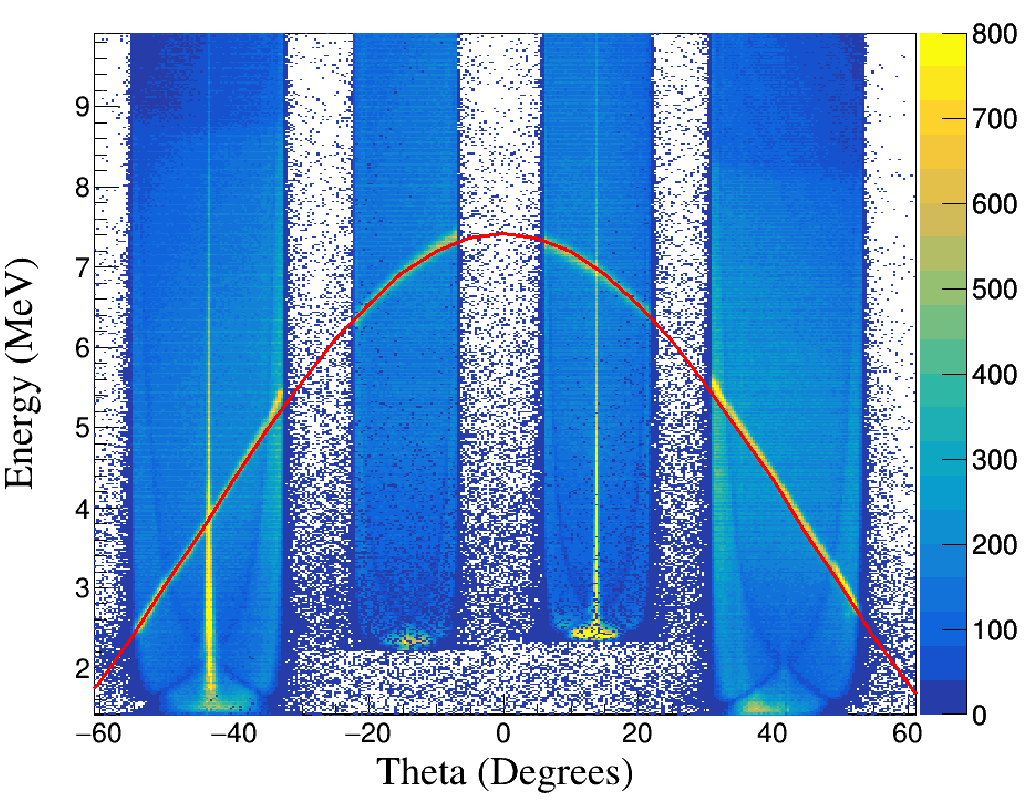}
\caption{Energy and position calibration of the PSD. The red curve shows the theoretical energy distribution of protons from $^{14}$N(p, p)$^{14}$N elastic scattering, while the yellow spots represent the measured energy deposition of scattered protons.}
\label{wv}
\end{figure}

The spatial calibration of the position-sensitive detectors was performed using an equally spaced calibration grating, while the energy response was calibrated with a standard $\alpha$ radioactive source in combination with the elastic scattering reactions $^{14}$N(p, p)$^{14}$N and $^{14}$N(d, d)$^{14}$N. Fig.~\ref{wv} presents the PSD energy-position calibration, where the horizontal axis represents the detector angle and the vertical axis represents the deposited particle energy. The red curve corresponds to the theoretical energy distribution of protons from $^{14}$N(p, p)$^{14}$N elastic scattering, while the bright yellow spots represent the energy deposition of scattered protons recorded in the PSD. As illustrated, the experimental data align closely with the theoretical curve, confirming the reliability of both the energy and spatial calibrations and providing a solid foundation for subsequent quantitative analyses.

\begin{figure}[htbp]
\includegraphics[width=8.5cm]{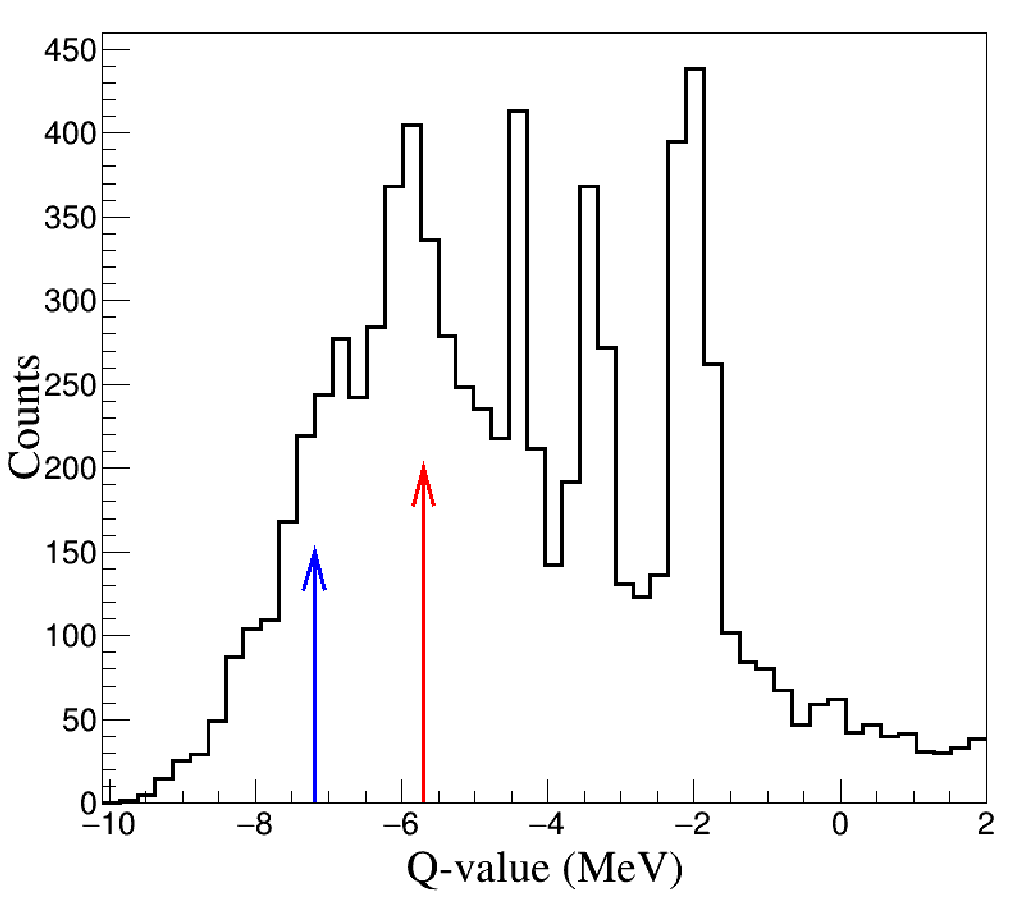}
\caption{Q-value spectrum of the $^{12}$C($^{14}$N,$\alpha d$)$^{20}$Ne reaction. The intersections of the red and blue solid lines with the Q-value axis, at -5.65 MeV and -7.28 MeV, correspond to the contributions from the ground state and the first excited state of the residual nucleus, respectively.}
\label{wvi}
\end{figure}

\begin{figure}[htbp]
\includegraphics[width=8.5cm]{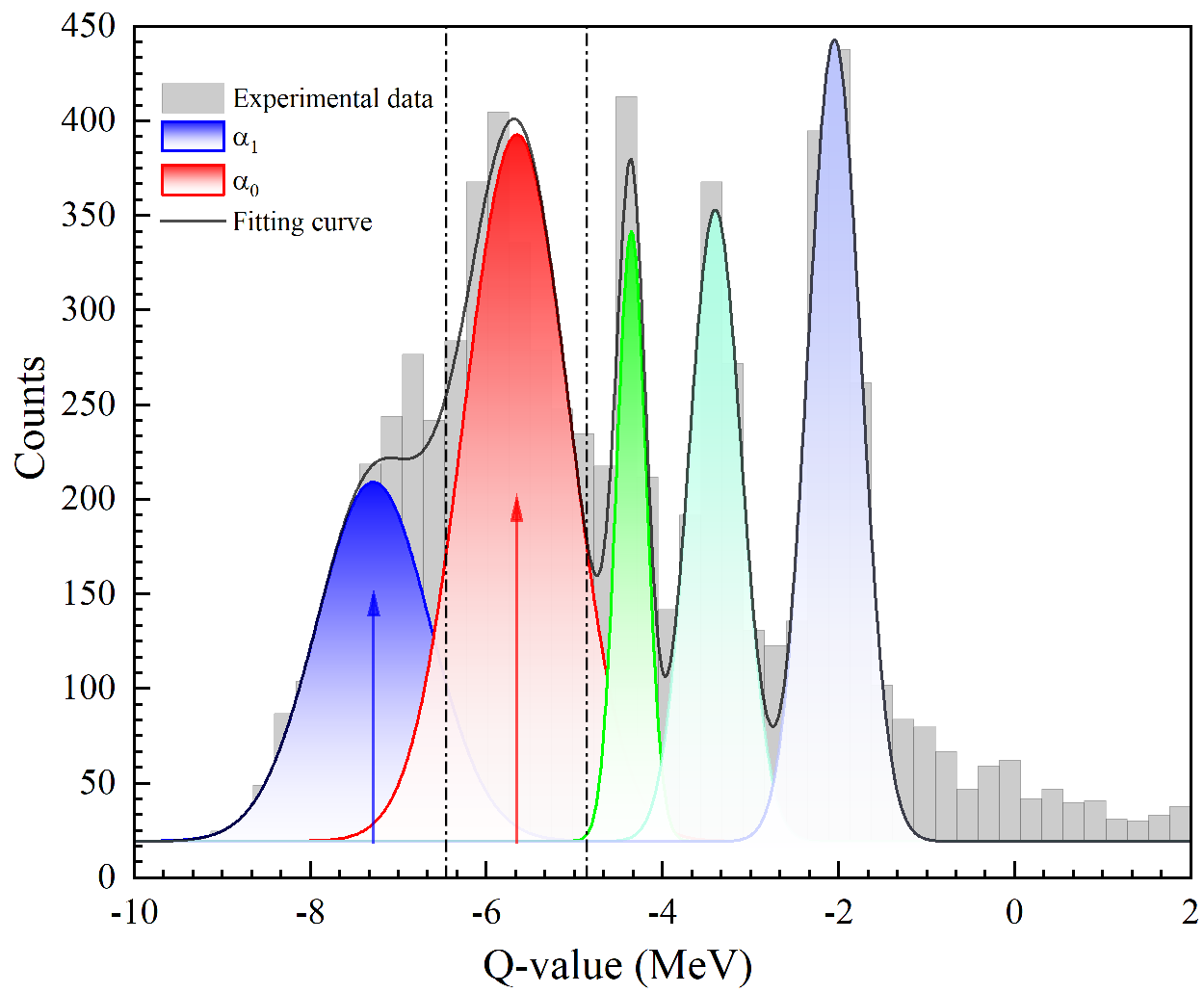}
\caption{Gaussian fitting of the Q-value spectrum for the $^{12}$C($^{14}$N,$\alpha d$)$^{20}$Ne reaction. The green area represents background events, while the red and blue regions correspond to the ground state and the first excited state, respectively.}
\label{wvx}
\end{figure}

The reaction $Q$ value is a fundamental observable that characterizes the energy balance of a nuclear reaction. Fig.~\ref{wvi} shows the reconstructed three-body $Q$-value spectrum for the $^{12}$C($^{14}$N,$\alpha d$)$^{20}$Ne reaction. The fitted centroid positions of the two dominant peaks, located at $-5.65$ MeV and $-7.28$ MeV, are consistent with the theoretical $Q$ values for the ground state ($\alpha_{0}$) and first excited state ($\alpha_{1}$) transitions of the residual nucleus, respectively. In addition, structures observed between approximately $-1$ MeV and $-5$ MeV mainly originate from competing reaction channels, such as the $^{12}$C($^{14}$N,$\alpha p$)$^{21}$Ne reaction and its excited states, reconstructed under the $\alpha+d+^{20}$Ne hypothesis.

As shown in Fig.~\ref{wvx}, the experimental $Q$-value spectrum is described by a multi-Gaussian fit, in which the green component represents the background contribution, while the red and blue components correspond to the $\alpha_{0}$ and $\alpha_{1}$ transitions, respectively. The multi-Gaussian function adopted in the fit is expressed as:
\begin{equation}
y = y_{0} + \frac{A}{w \sqrt{\pi / 2}} \exp \left[ -2 \left( \frac{x - x_{c}}{w} \right)^{2} \right]
\label{e10i}
\end{equation}
where $y_{0}$ is the baseline offset, $A$ is the total area under the peak, $x_{c}$ is the peak centroid position, and $w$ denotes the width parameter. The fit reproduces the experimental spectrum well, yielding a coefficient of determination of $R^{2} = 0.94$. The fitted width parameters for the $\alpha_{0}$ and $\alpha_{1}$ peaks are $w_{\alpha_{0}} = 1.20 \pm 0.10~\mathrm{MeV}$ and $w_{\alpha_{1}} = 1.29 \pm 0.15~\mathrm{MeV}$, corresponding to standard deviations of $\sigma_{\alpha_{0}} = 0.60 \pm 0.05~\mathrm{MeV}$ and $\sigma_{\alpha_{1}} = 0.64 \pm 0.08~\mathrm{MeV}$, respectively. The derived full widths at half maximum (FWHM) are $\mathrm{FWHM}_{\alpha_{0}} = 1.41 \pm 0.12~\mathrm{MeV}$ and $\mathrm{FWHM}_{\alpha_{1}} = 1.52 \pm 0.18~\mathrm{MeV}$, which remain consistent with each other within experimental uncertainties. Owing to the finite experimental energy resolution, a significant overlap occurs between the broad $\alpha_{0}$ and $\alpha_{1}$ peaks. In their overlap region, the superimposed statistical contributions of both peaks elevate the experimental counts, leading to an apparent shift in peak position and spectral shape distortion. The FWHM value of approximately $1.4\text{--}1.5~\mathrm{MeV}$ extracted from the fit reflects the total width of the reconstructed three-body $Q$-value spectrum, rather than the intrinsic energy resolution of the detector system itself.

\begin{equation}
P(A|B)=\frac{P(AB)}{P(B)}=\frac{P(B|A)\cdot P(A)}{P(B)}
\label{e9}
\end{equation}%

To focus on the $\alpha_{0}$ channel, the present analysis adopts a $Q$-value gate of $-5.65 \pm 0.8~\mathrm{MeV}$, which serves to reduce the influence of competing reaction channels while ensuring sufficient count statistics. To quantitatively evaluate the peak overlap and background effects within this candidate region, the Bayesian method introduced in Eq.~(\ref{e9}) is used to estimate the statistical contribution of the overlapping $\alpha_{1}$ events and other competing processes. The resulting contamination is estimated to be less than 10\%, thereby reducing the influence of background events as much as possible.

It should be emphasized that the identification of the $\alpha_{0}$ channel is not based on the Bayesian analysis alone. Instead, the final event selection is achieved through the combined application of multiple selection criteria, including particle identification, detector-geometry constraints, reconstructed three-body $Q$ values, and quasi-free kinematic conditions described by Eqs.~(\ref{e6})--(\ref{e8}), thereby effectively suppressing contributions from competing reaction channels.

\subsection{Selection of Quasi-Free Reaction Events}

\begin{figure}[tbp]
\includegraphics[width=8.5cm]{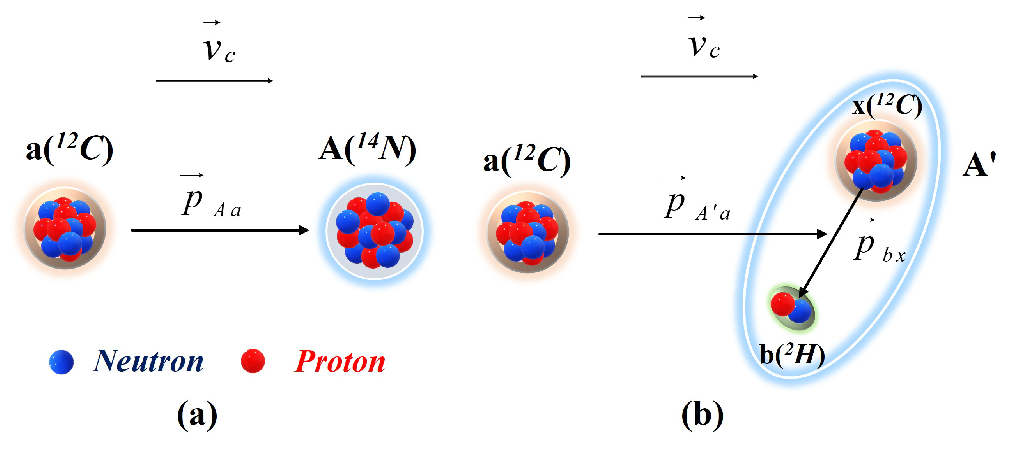}
\caption{Schematic illustration of the intermediate breakup process $^{12}$C + $^{14}$N $\rightarrow$ $^{12}$C + $^{12}$C + $^{2}$H. The $\vec{v}_{c}$ denotes the velocity in the center-of-mass frame.}
\label{wvii}
\end{figure}

Fig.~\ref{wvii} illustrates a possible intermediate breakup of the Trojan horse nucleus during the reaction, specifically the process $^{12}$C + $^{14}$N $\rightarrow$ $^{12}$C + $^{12}$C + $^{2}$H. By combining the physical quantities depicted in Fig.~\ref{wvii} with the principles of energy and momentum conservation, Eqs.~(\ref{e6})-(\ref{e8}) can be derived ~\cite%
{Wen2008,Wen2011}.

\begin{equation}
\frac{p^{2}_{A'a}}{2\mu_{A'a}}+\frac{p^{2}_{bx}}{2\mu_{bx}}=\frac{p^{2}_{Aa}}{2\mu_{Aa}}+Q'
\label{e6}
\end{equation}%
\begin{equation}
Q'=(m_{A}-m_{b}-m_{x})c^{2}
\label{e7}
\end{equation}%
\begin{equation}
\overrightarrow{p_{bx}}+\frac{m_{b}}{m_{A'}}\overrightarrow{p_{A'a}}+m_{b}\overrightarrow{v_{c}}=\overrightarrow{p_{b}}
\label{e8}
\end{equation}%

In Eq.~(\ref{e8}), $\overrightarrow{p_{b}}$ denotes the momentum of b ($^{2}$H) in the laboratory frame, while particle a ($^{12}$C) represents the stationary target nucleus in the experiment. By applying Eqs.~(\ref{e6})-(\ref{e8}), the momentum of b ($^{2}$H) obtained from the experimental data can be used to calculate either $\textit{p}_{A'a}$ or $\textit{p}_{bx}$. If the calculated value of $\textit{p}_{A'a}$ or $\textit{p}_{bx}$ is a positive real number, it indicates the occurrence of the intermediate breakup process $^{12}$C + $^{14}$N $\rightarrow$ $^{12}$C + $^{12}$C + $^{2}$H. This condition was adopted as one of the criteria for selecting quasi-free events.

\begin{figure}[htbp]
\includegraphics[width=8.5cm]{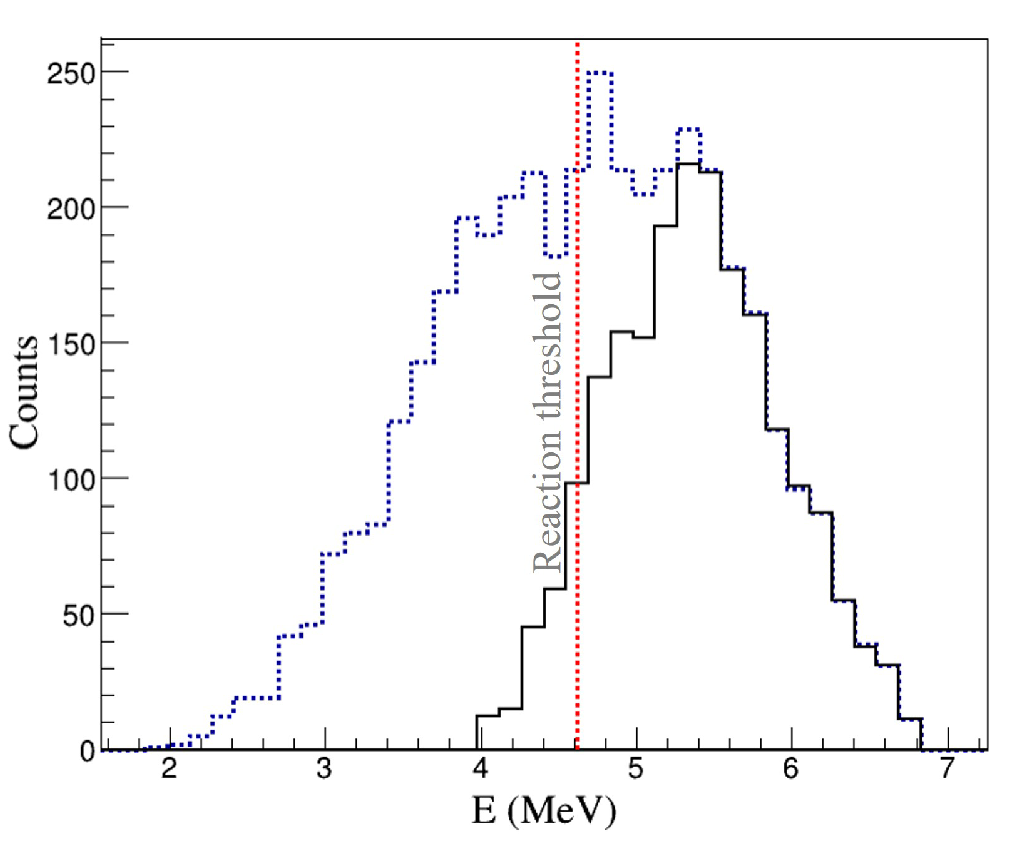}
\caption{Relative energy spectrum of $^{20}$Ne and $\alpha$ particle with the intermediate breakup process constraint applied. The blue dashed and black solid histograms represent the spectra before and after applying the constraint, respectively.}
\label{wviii}
\end{figure}

In the application of the THM, quasi-free events are typically selected by imposing constraints on the spectator momentum. However, in the study of the $^{9}$Be(p,$\alpha$)$^{6}$Li reaction [22, 23], it was found that restricting only the spectator momentum was insufficient to effectively exclude contributions from cascade processes, which could lead to an enhancement of resonance structures. To address this issue, subsequent analyses incorporated the intermediate breakup process based on energy and momentum conservation relations, thereby imposing both kinematic and dynamical constraints on quasi-free events.

This approach improved the physical reliability of event selection and successfully yielded reliable bare-nucleus cross-section data. Therefore, we introduce the intermediate breakup process as an additional selection condition to further strengthen the physical constraints on event selection and enhance the reliability of quasi-free event identification.

Fig.~\ref{wviii} presents the relative energy spectra of $^{20}\text{Ne}$ and $\alpha$ particles before and after applying the intermediate breakup (BU) kinematics constraints. It is worth emphasizing the sequence of the event selection procedure: prior to constructing these relative energy spectra, a $Q$-value gate of $-5.65 \pm 0.8\text{ MeV}$ was already imposed on all candidate events. Consequently, the prominent background structures observed at higher $Q$ values in Fig.~\ref{wvx} (e.g., around $-2$, $-3.5$, and $-4.4\text{ MeV}$, which originate from competing channels such as $p+\alpha+^{21}\text{Ne}$) were suppressed as far as possible before entering the analysis shown in Fig.~\ref{wviii}.

In Fig.~\ref{wviii}, the blue dashed and black solid histograms represent the $Q$-gated spectra before and after applying the intermediate breakup constraints, respectively, with the red dashed line marking the reaction threshold at $4.62\text{ MeV}$. As clearly demonstrated in the figure, imposing the intermediate breakup condition further suppresses residual background events near the reaction threshold, thereby further reducing background contamination for the target $^{12}\text{C}(^{12}\text{C},\alpha_0)^{20}\text{Ne}$ channel.

\subsection{Analysis of the Deuteron Momentum Distribution}

\begin{figure}[htbp]
\includegraphics[width=8.5cm]{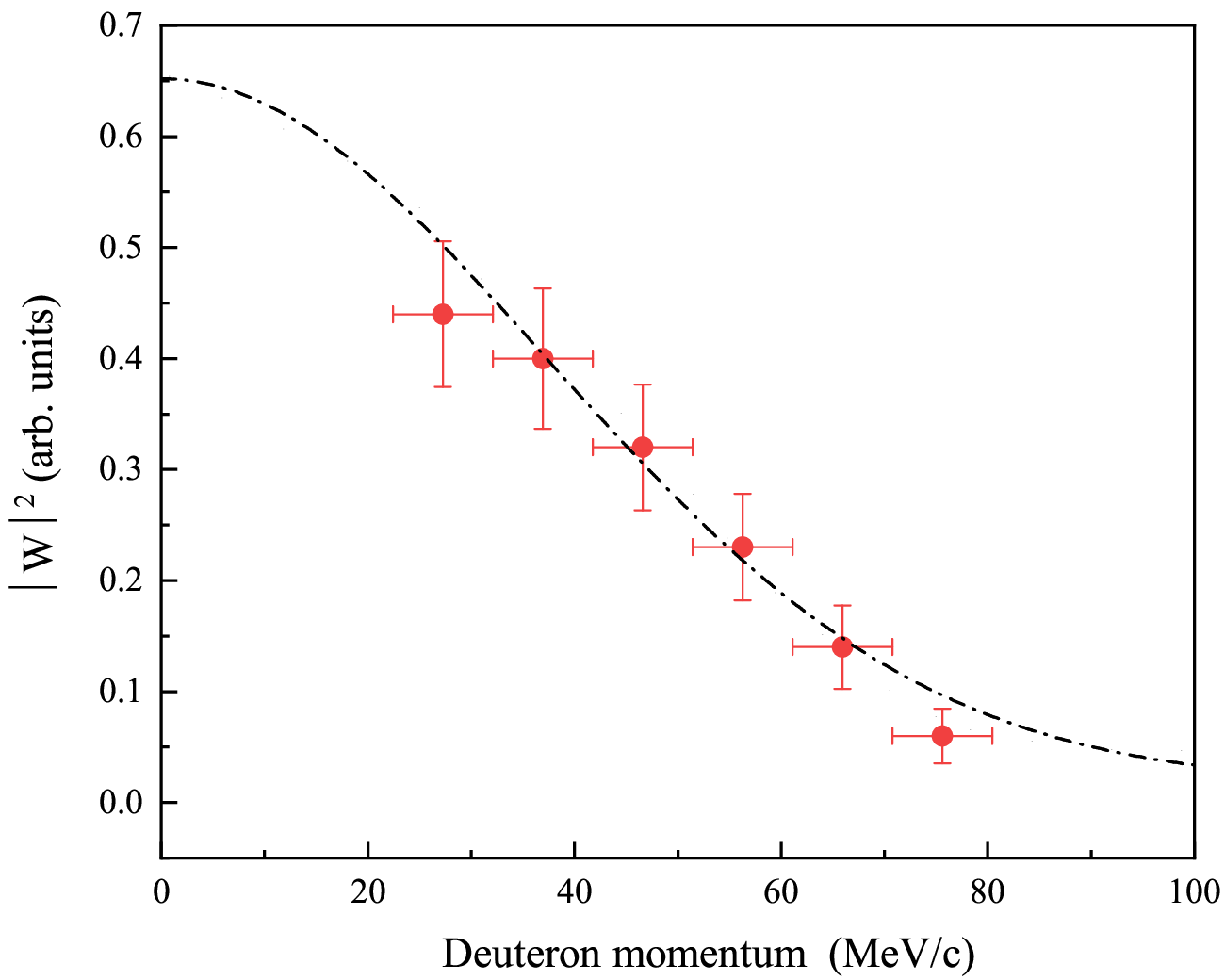}
\caption{Comparison between the experimental momentum distribution of the spectator deuteron (red circles) and the theoretical distribution (black dashed line). The error bars represent statistical uncertainties.}
\label{wxi}
\end{figure}

The momentum distribution of the spectator deuteron is a highly sensitive observable for investigating the reaction mechanism, as it provides a direct means to validate the selection of quasi-free events. According to the quasi-free reaction mechanism, the momentum distribution of the spectator particle is expected to reproduce its internal momentum distribution within the parent nucleus.

For the extraction of the astrophysical $S^{*}(E)$ factor in the center-of-mass energy range of 0.5--2.0 MeV, the spectator-deuteron momentum distribution was reconstructed to examine the quasi-free reaction mechanism. A relatively narrow $^{20}$Ne--$\alpha$ relative-energy window of 50--100 keV was adopted so that the quantity $\textit {p}_{l}\frac{d\sigma_l}{d\Omega}$ could be regarded as approximately constant within each energy interval. The center of the selected energy window was varied across the investigated energy range to maintain sufficient statistics while minimizing the influence of energy broadening. In addition, the event sample was further constrained by the intermediate-breakup selection and the kinematic conditions described by Eqs.~(\ref{e6})--(\ref{e8}).

The momentum distribution of the spectator deuteron serves as an important criterion for verifying the validity of the quasi-free reaction mechanism. Therefore, the experimentally reconstructed momentum distribution was compared with theoretical calculations. Preliminary candidate events for the reaction channel of interest were selected using $\Delta E\text{--}E$ particle identification, experimental geometric constraints, and three-body kinematic $Q$-value reconstruction. Subsequently, kinematic constraints from the intermediate breakup process were applied to further suppress contributions from competing reaction mechanisms. On this basis, the agreement between the experimental spectator deuteron momentum distribution and the theoretical quasi-free distribution was analyzed to determine the optimal quasi-free event selection region for subsequent analysis.

In Fig.~\ref{wxi}, the reconstructed spectator-deuteron momentum distribution for the $^{12}$C($^{14}$N,$\alpha d$)$^{20}$Ne reaction is presented as red solid circles with a momentum-bin width of 10 MeV/$c$. The error bars represent the statistical uncertainties. The black dashed curve shows the theoretical momentum distribution calculated using a Woods--Saxon potential with standard geometrical parameters. The potential depth was slightly adjusted to reproduce the experimental ground-state binding energy of 10.27 MeV for the $^{12}$C--d cluster configuration in $^{14}$N. The experimental and theoretical distributions were normalized to the same integral for comparison.

The absence of experimental data points below approximately 30 MeV/$c$ is mainly related to the experimental detection threshold. Low-momentum spectator deuterons suffer from significant energy losses when passing through the blocking foil and the dead layers of the position-sensitive detector, resulting in reduced detection efficiency in this momentum region. In addition, electronic noise suppression and event-quality requirements applied during the data analysis further reduce the available statistics at low momenta. Therefore, the present momentum distribution is limited to the experimentally accessible region.

For the two experimental setups, the reconstructed spectator-deuteron momentum distribution exhibits a trend consistent with the theoretical quasi-free distribution in the dominant region of 30--80 MeV/$c$, providing strong support for the adopted spectator-momentum selection. Consequently, the momentum window of $30 < p_d < 80$ MeV/$c$ was adopted for selecting quasi-free events, and the events within this window were utilized for the subsequent $S^*(E)$ factor analysis.

\section{Extraction and Analysis of the Astrophysical $S^*(E)$ Factor}
\subsection{The Astrophysical $\textit{S*(E)}$ Factor}

\begin{figure*}[htbp]
\centering
\includegraphics[width=\linewidth]{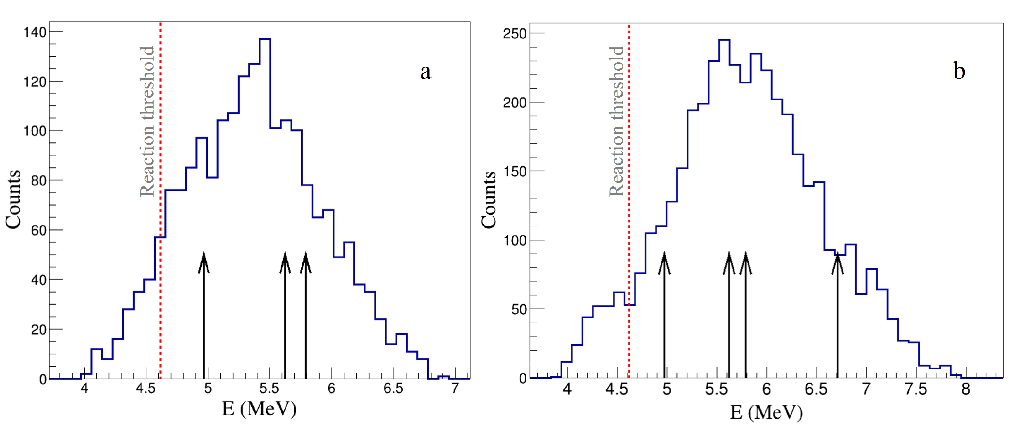}
\caption{Relative energy spectrum of $^{20}$Ne and $\alpha$ particle. (a) Spectator deuteron with a central angle of 15$^\circ$; (b) Spectator deuteron with a central angle of 0$^\circ$. Resonance structures at 4.967, 5.621, 5.788, and 6.706 MeV are indicated by arrows. These resonance structures correspond to known $^{12}$C+$^{12}$C resonances in the center-of-mass energy range $E_{\mathrm{c.m.}}$ = $0-3$ MeV.}
\label{wvix}
\end{figure*}

Fig.~\ref{wvix} presents the reconstructed relative-energy ($^{20}\text{Ne}+\alpha$) spectra obtained after applying the complete event-selection chain, which integrates particle identification, detector geometrical bounds, three-body $Q$-value gating, intermediate-breakup constraints, and the spectator-momentum cut ($30 < p < 80\text{ MeV}/c$). Panels (a) and (b) represent the data measured with spectator-deuteron detection centered at $15^\circ$ and $0^\circ$, respectively. Due to the limited experimental energy resolution, potential narrow resonances undergo energy broadening during the reconstruction process and overlap with adjacent structures, forming a broad distribution alongside potential non-resonant continuum contributions. Known resonance positions are indicated by vertical arrows to serve as spectroscopic reference points. Rather than forcing an unconstrained multi-component decomposition into individual resonances, which would introduce substantial parameter ambiguities ,the present analysis focuses on extracting the robust overall energy dependence of the astrophysical $S^*(E)$ factor across the Gamow window.

In the center-of-mass system, the astrophysical $\textit{S*(E)}$ factor of the $^{12}$C+$^{12}$C fusion reaction is defined by Eq.~(\ref{e10}):

\begin{equation}
\textit{S*(E)}=\sigma(E)E\mathrm{exp}(\frac{87.21}{\sqrt{E}}+0.46E)
\label{e10}
\end{equation}%

By combining multiple selection criteria, including particle identification, angular constraints imposed by the experimental setup, kinematic conditions associated with the intermediate breakup mechanism, and additional Q-value gating the contribution of competing reaction channels and background events is significantly reduced. Due to the effects of nucleon-nucleon and Coulomb interactions, the present analysis is performed within the framework of the DWBA.

\begin{figure}[htbp]
\includegraphics[width=8.5cm]{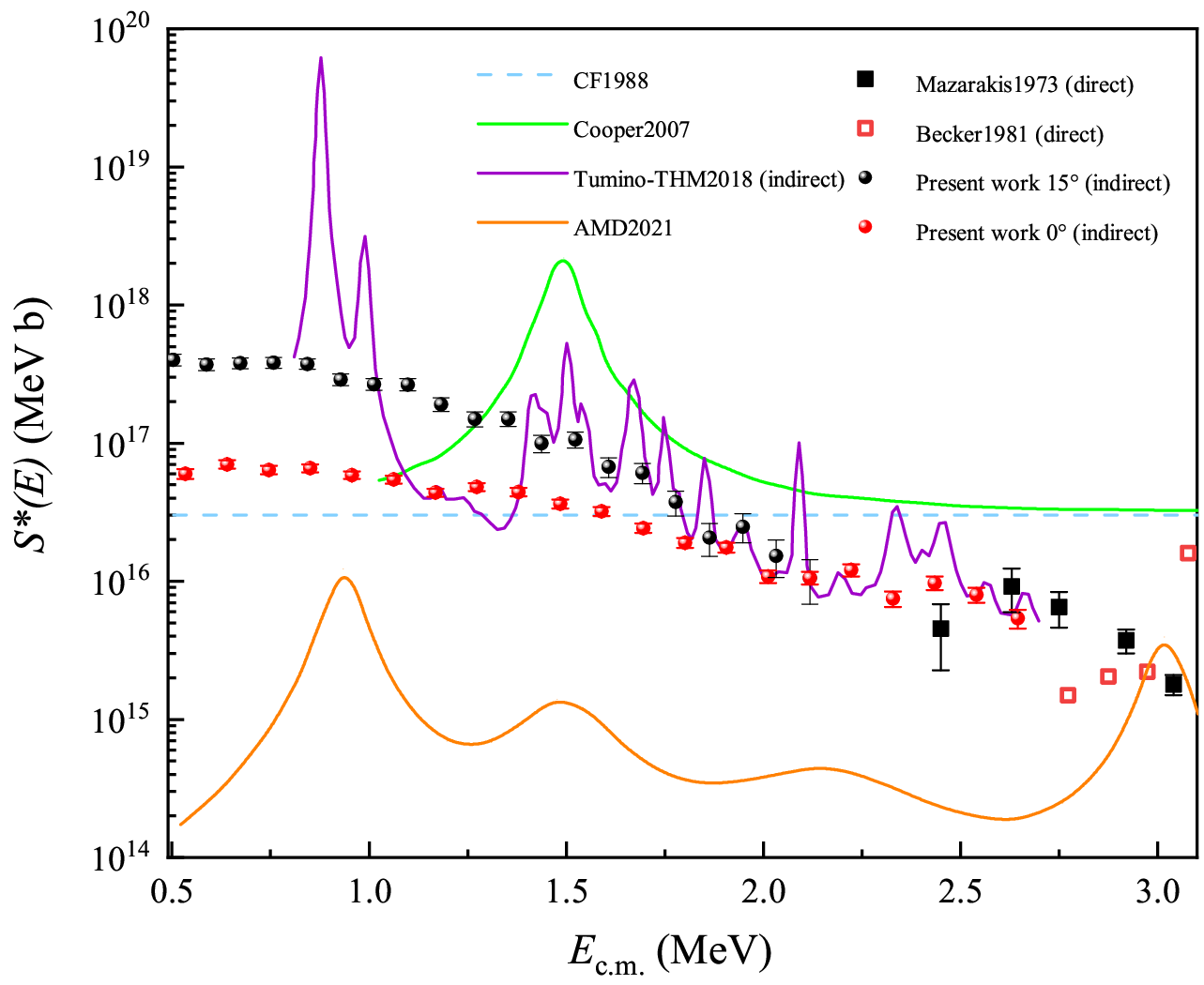}
\caption{Experimentally extracted $\textit{S*(E)}$ factor for the two-body reaction $^{12}$C($^{12}$C,$\alpha_{0}$)$^{20}$Ne after quasi-free event selection, compared with existing experimental data and theoretical predictions. Some of the referenced datasets do not explicitly isolate the $\alpha_{0}$ branch but instead include contributions from the inclusive  channel; these are retained to facilitate comparison of the overall energy dependence. The reference data are taken from Refs. \cite{Rolfs1988,Becker1981,Tumino2018,Cooper2009,Taniguchi2021,Mazarakis1973}. The error bars include statistical uncertainties. The uncertainty in the center-of-mass energy is approximately $\sigma_{E_{\rm c.m.}} = 0.3$ MeV (FWHM = 0.70 MeV).}
\label{wxii}
\end{figure}

The astrophysical $\textit{S*(E)}$ factor was calculated using Eqs.~(\ref{e5}) and (\ref{e10}). The experimentally extracted $\textit{S*(E)}$ factors were then normalized to the existing measurements and systematically compared with other experimental data and theoretical model calculations ~\cite%
{Rolfs1988,Becker1981,Tumino2018,Cooper2009,Taniguchi2021,Mazarakis1973}. Fig.~\ref{wxii} shows two sets of $\textit{S*(E)}$ factors for the $^{12}$C($^{12}$C,$\alpha_{0}$)$^{20}$Ne reaction obtained after the quasi-free event selection, represented by black and red solid circles, respectively.

Both data sets were normalized to the data of Tumino 2018 \cite{Tumino2018} and the energy corrected data of Mazarakis 1973 \cite{Mazarakis1973}, which includes the standard $+100\text{ keV}$ energy shift correction as recommended by Barnes et al. \cite{Barnes1985}. The error bars include statistical uncertainties. The black solid circles correspond to spectator deuteron angles in the range of 7$^\circ$$\sim$23$^\circ$, which is comparable to the experimental condition in Tumino 2018 ~\cite%
{Tumino2018}, while the red solid circles correspond to the range of -8$^\circ$$\sim$8$^\circ$.

As shown in Fig.~\ref{wxii}, the extracted $S^{*}(E)$ factors from both data sets are objectively presented. From the present $S^*(E)$ factor data, the overall feature of the $^{12}$C+$^{12}$C reaction in the low-energy region can be viewed as a series of resonance structures superimposed on a relatively flat non-resonant continuum. As the energy decreases, the cumulative superposition of these low-energy resonances macroscopically drives the overall $S^*(E)$ factor to exhibit an increasing physical trend from the high-energy to the low-energy region. Without resolving the fine resonance structures, potential narrow resonances may undergo energy broadening and overlap with one another. In contrast, the red data points display a flatter energy dependence and remain lower than the black data points in the low-energy region. In addition, the present experimental data show possible bump-like structures around 0.8 and 1.5 MeV, which may be associated with the superposition of broadened resonance contributions.

The uncertainty in the center-of-mass energy is approximately $\sigma_{E_{\rm c.m.}} = 0.3$ MeV (FWHM = 0.70 MeV). The uncertainty in the center-of-mass energy is mainly determined by energy broadening and geometrical uncertainties in the detection angles. In particular, the angular uncertainty is jointly constrained by the position resolution of the PSD detector, the beam-spot size, and the distance between the detector sensitive area and the target position, which determines the geometrical solid angle and angular acceptance. It is also affected by the accuracy of the detector angular calibration. The energy uncertainty mainly originates from the intrinsic energy resolutions of the PSD and SSSD detectors, uncertainties in energy calibration, uncertainties in the energy-loss correction for the detector dead layers, and energy straggling of the light particles in the beam-stopper foil, with the nonuniformity in the foil thickness being an important contributing factor.

Since $^{20}\mathrm{Ne}$ is not directly detected as the third outgoing particle, its momentum and energy are reconstructed entirely from the three-body kinematics through energy and momentum conservation, based on the measured kinematic information of the two light particles ($p$ and $\alpha$). Therefore, the energy and angular resolutions of the directly measured light particles are propagated through the differential kinematic relations and accumulated in the reconstructed energy and angular quantities of $^{20}\mathrm{Ne}$, thereby contributing collectively to the broadening of the total center-of-mass energy.

The present data do not allow a reliable separation of individual resonance contributions. Consequently, the present analysis focuses on the overall energy dependence rather than the fine structure of individual resonances. Multiple selection criteria including particle identification, angular constraints, three-body $Q$-value restrictions, intermediate-breakup channel selection, and spectator-deuteron momentum cuts were applied to both experimental setups, which effectively reduced the contribution of non-quasi-free events. The discrepancy observed between the two data sets in the low-energy region may originate from a combination of several factors: experimental acceptance effects, finite detector resolution, and potential variations in the relative contributions of different reaction mechanisms.

\subsection{Monte Carlo Simulations and Background Analysis}

\begin{figure}[htbp]
\includegraphics[width=8.5cm]{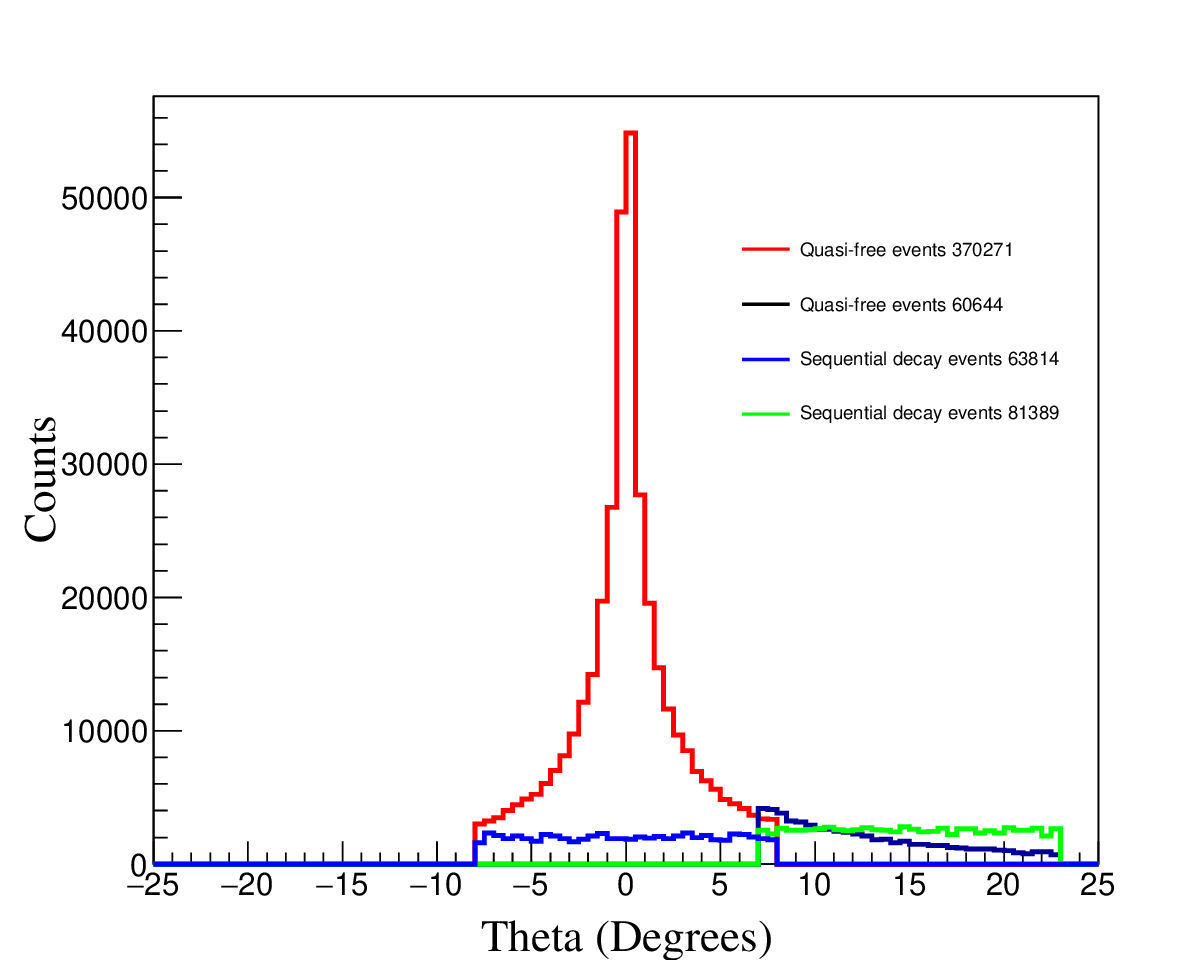}
\caption{ Simulated comparison of event counts from quasi-free and sequential decay processes.}
\label{wxiii}
\end{figure}

Fig.~\ref{wxiii} compares the simulated event yields for the quasi-free and sequential decay mechanisms under the two experimental configurations. To evaluate the corresponding detection acceptances, $5\times10^{6}$ events were generated for each reaction mechanism and propagated through the experimental geometry. The red and black curves represent the accepted quasi-free events for spectator central angles of $0^\circ$ and $15^\circ$, respectively, while the blue and green curves correspond to the accepted sequential-decay events for the two configurations. The simulation indicates that the $0^\circ$ configuration provides a higher acceptance for quasi-free events, whereas the relative contribution from sequential decay is reduced under the same selection criteria.

Within the THM framework, the quasi-free reaction mechanism is expected to dominate when the spectator particle is emitted close to the beam direction with low momentum. This expectation is supported by the present Monte Carlo simulations and by the measured spectator angular distribution. Consequently, the $0^\circ$ configuration is expected to enhance the selection efficiency for quasi-free events. If competing reaction mechanisms are not sufficiently suppressed during the event-selection procedure, their residual contributions may affect the reconstructed observables and consequently influence the extracted astrophysical $S^{*}(E)$ factor. Therefore, particle identification, three-body $Q$-value reconstruction, intermediate-breakup selection, and spectator-momentum constraints were applied simultaneously to minimize the contribution from competing processes.

\subsection{Discussion of Experimental Results}

The differences observed in the results between the two experimental setups can be mainly attributed to the following factors:

(1) The target reaction channel ($\alpha_0$ channel) is affected by the $\alpha_1$ channel as well as neighboring reaction channels. Candidate events for the target reaction channel were preliminarily selected using $\Delta E\text{--}E$ particle identification, experimental geometric constraints, and three-body kinematic $Q$-value reconstruction. Subsequently, kinematic constraints from the intermediate breakup process were applied to further suppress contributions from competing reaction mechanisms. On this basis, the influence of non-quasi-free events was minimized as much as possible by reconstructing the spectator-deuteron momentum distribution and comparing it with the theoretical Woods--Saxon distribution.

(2) As shown in Figs.~\ref{wvix}(a) and \ref{wvix}(b), an overall structure formed by the superposition of low-energy resonances can still be observed in the center-of-mass energy range of approximately 0--3 MeV; however, the resolution is insufficient to separate adjacent narrow resonances from one another. Fig.~\ref{wvix} demonstrates the objective existence of resonance contributions in this energy region. Although the $S^*(E)$ factors extracted on this basis exhibit certain numerical discrepancies, their overall energy dependence trends remain consistent.

(3) Monte Carlo simulation results indicate that the two experimental setups differ in their acceptance efficiencies for the quasi-free reaction process. Under the present experimental conditions, due to limited statistics, detector resolution, and the presence of competing reaction mechanisms, the experimental analysis cannot completely eliminate all non-quasi-free contributions. Therefore, the $S^*(E)$ factors extracted in this work are mainly intended to study the overall trend in the low-energy region, rather than to provide a precise determination of individual resonance parameters.

In summary, considering the aforementioned experimental constraints and acceptance effects, the primary physical conclusion that can be drawn from the present study is that the low-energy region can be characterized as a series of possible resonance structures superimposed on a relatively flat non-resonant continuum. As the energy decreases, the cumulative superposition of these low-energy resonances macroscopically drives the overall $S^*(E)$ factor to exhibit an increasing physical trend.

\section{Summary and Discussion}

In this work, the THM was applied to perform an indirect measurement of the $^{12}$C($^{12}$C,$\alpha_{0}$)$^{20}$Ne reaction, which plays a crucial role in stellar evolution and supernova explosions. Since direct measurements in the astrophysical energy region are extremely challenging due to strong Coulomb barrier suppression, the three-body reaction $^{12}$C($^{14}$N,$\alpha d$)$^{20}$Ne was investigated to extract the astrophysical $S^*(E)$ factor, employing $^{14}$N as the Trojan horse nucleus.

Guided by the quasi-free mechanism, which indicates that spectator particles emerge primarily in the forward-angle region, the experimental geometry was selected to explore the kinematic phase space where quasi-free contributions are expected to be significant. The experiment employed a $\Delta E-E$ silicon telescope detection system with two configurations designed to systematically cover distinct forward angular ranges, including measurements near 0$^\circ$. Furthermore, the intermediate breakup process $^{12}$C+$^{14}$N $\rightarrow$ $^{12}$C+$^{12}$C+$^{2}$H was introduced as an additional physical constraint to investigate the influence of spectator kinematic selection.

Based on the distorted-wave Born approximation (DWBA) analysis, two sets of astrophysical $S^{*}(E)$ factors for the $^{12}$C($^{12}$C,$\alpha_{0}$)$^{20}$Ne reaction were extracted and normalized to the available direct measurements of Mazarakis \textit{et al.} and the previous Trojan Horse Method (THM) results reported by Tumino \textit{et al.} Within the astrophysical energy range of 0.5--2 MeV, the $S^{*}(E)$ factor exhibits an overall increasing trend with decreasing energy. The results obtained with the $0^\circ$ configuration display a flatter energy dependence, with generally lower $S^{*}(E)$ values in the low-energy region. The difference observed between the two data sets in the low-energy region may originate from a combination of several factors: experimental acceptance effects, finite detector resolution, and potential variations in the relative contributions of different reaction mechanisms.

Although the present experiment does not allow a precise determination of individual resonance structures, this comparative study reveals the overall energy dependence of the $S^*(E)$ factor in the 0.5--2.0 MeV region, an issue central to the ongoing debate surrounding the $^{12}$C+$^{12}$C fusion reaction. From the present $S^*(E)$ factor data, the overall feature of the $^{12}$C+$^{12}$C reaction in the low-energy region can be viewed as a series of resonance structures superimposed on a relatively flat non-resonant continuum. As the energy decreases, the cumulative superposition of these low-energy resonances macroscopically drives the overall $S^*(E)$ factor to exhibit an increasing physical trend. Moreover, potential narrow resonances may undergo energy broadening and overlap with one another, thereby manifesting experimentally as a flatter energy dependence.

By systematically examining the experimental design, quasi-free event selection strategies, and interpretations of the underlying physical mechanisms under two different experimental configurations, these results provide a useful reference for understanding stellar nucleosynthesis and guiding future high-precision measurements. Ultimately, through the combination of direct measurements, indirect approaches, and theoretical models, the underlying mechanisms of this critical reaction can be further constrained.

\section*{Acknowledgments}
The authors sincerely thank Prof. Shu-Hua Zhou from the CIAE for his valuable advice and guidance throughout this work. The authors would like to express their gratitude to the Nuclear Reaction Group at the CIAE for their kind assistance during the experimental preparation and measurement. Special thanks are extended to the staff of the HI-13 tandem accelerator for their strong support in ensuring the smooth completion of the experiment.

This work was supported by the National Natural Science Foundation of China (Grants Nos. 12075031, 12275360, 12475115, and 11935001), the Natural Science Foundation of Beijing Municipality (1222022), and the Doctoral Foundation of the Anhui Provincial Department of Education (2025AHGXZK50097).


\end{document}